# Statistical Model of Time-varying Backscatter Power of Monostatic RF Sensing Channels in Urban Canyons

Dmitry Chizhik, *Fellow, IEEE*, Jakub Sapis, *Member, IEEE,* John Drogo, Abhishek Adhikari, Manuel Almendra, Jinfeng Du, *Senior Member, IEEE,* Reinaldo A. Valenzuela, *Fellow, IEEE*, Gil Zussman, *Fellow, IEEE*, Mauricio Rodriguez, *Senior Member, IEEE*, and Rodolfo Feick, Life *Senior Member, IEEE*

***Abstract*— We present a measurement-based statistical model for the backscatter power ratio of monostatic RF sensing in urban canyons with moving clutter, suitable for large-scale system level performance evaluation of RF sensing in 6G networks. A narrowband (CW) 140 GHz sounder used a monostatic radar arrangement with an omnidirectional transmit antenna illuminating streets and a spinning horn 2º receive antenna offset vertically (less than 1 m away) collecting backscattered power as a function of azimuth and time below building height in Manhattan and Valparaiso, Chile. A concise outdoor deterministic model of average backscattered power dependent on distance to nearest building-wall reproduces observations with 3.3 dB RMS error or better. Distribution of power variation in azimuth around this average is reproduced within 0.5 dB by a random azimuth spectrum with a lognormal distribution. Temporal fluctuations for various antenna aims and locations were found to be well modeled by a Rician distribution, with lognormally distributed K-factor, with 0.47-0.73 correlation coefficient to backscatter power deviation from mean. The statistical model does not require a detailed environmental description, aiming to reproduce backscatter clutter statistics (as opposed to a deterministic response) faithfully and efficiently, essential for large-scale system-level performance evaluation.**

## I. INTRODUCTION

There is an increasing interest [1][2][3][4] in the use of communication signals for sensing, often termed Joint/Integrated Communications and Sensing (JCAS or ISAC*)*. The new functionality is essentially that of radar, aiming at detecting and possibly characterizing some aspects of the environment. Often the goal is to detect and localize a static or moving object, termed "target", such as a person, vehicle, robot, or drone, in the presence of the rest of the environment, the response to which is termed "clutter". Some of the applications of outdoor RF sensing being discussed [5] include detection of people and vehicles in public spaces for crowd/traffic monitoring or public safety.

Clutter echoes are often stationary, e.g, buildings outdoors, walls and furniture indoors but may also vary in time as the environment includes moving objects such as people, vehicles, as well as vegetation and other objects that sway with the wind. Sensing scenarios of interest include monostatic, with co-located transmitter/receiver measuring backscatter, e.g., a single base station or terminal, as well as bi-static where transmitter and receiver are separated, e.g., signal traveling from one base station to another, scattering along the way. "Monostatic" arrangement discussed here includes quasi-monostatic, where transmitter and receiver are only separated enough (say within 1 m) to achieve adequate isolation. Various use cases for ISAC are under consideration by the 3GPP [5].

Algorithms for detection, localization and, possibly, classification, of objects need to be tested in realistic scenarios, requiring representative models of both clutter and target. For example, detection of a moving target (e.g. person or vehicle) in the presence of static or slowly varying clutter can be done digitally by subtracting static signals from a time sequence of backscatter measurements. An echo from a moving target might thus be detectable provided the receiver has enough dynamic range and is not saturated by the static clutter return. The detection performance will be critically dependent on the Signal to Background power ratio. Determination of adequate dynamic range requirements would thus be dictated by strengths of target and clutter echoes. Since target echo fluctuation is often the key aspect allowing distinction of target echo from background clutter, temporal variation in clutter returns, may be particularly confounding in attempting to detect and localize a target. A desirable background clutter model must thus provide its strength, distribution in delay and angle as well as temporal variation. It is desirable for the model to also be easily implementable.

The standard model recommendation 3GPP TR 38.901 [9] has recently been extended from communication-only channel model to now include sensing channels as well. The background clutter model for base station-based monostatic sensing in [9] prescribes a heuristic procedure to generate the background channel experienced by the Transmit-Receive

Dmitry Chizhik, Jakub Sapis, Jinfeng Du and Reinaldo A. Valenzuela are with Nokia Bell Labs, Murray Hill, NJ 07974, USA (dmitry.chizhik@nokia-bell-labs.com, jakub.sapis@nokia-bell-labs.com, jinfeng.du@nokia-bell-labs.com, reinaldo.valenzuela@nokia-bell-labs.com), John Drogo, Abhishek Adhikari and Gil Zussman, are with Columbia University, New York, NY (j.drogo@columbia.edu,abhishek.adhikari@columbia.edu, gil.zussman@columbia.edu), Manuel Almendra and Mauricio Rodgriguez are with Pontificia Universidad Católica de Valparaíso, Valparaíso, Chile (manuel.almendra.v@mail.pucv.cl, mauricio.rodriguez.g@pucv.cl), and Rodolfo Feick is with Universidad Técnica Federico Santa María, Valparaíso, Chile (rodolfo.feick@gmail.com). (*Corresponding author: Dmitry Chizhik)* point (TRP), e.g. a sensing base station: placing a set of 3 “reference points (RP)” on a circle around the TRP. The background clutter channel is then generated by treating the TRP-RP channel as a 1-way communication channel, with characteristics such as angle spread assigned using values specified for the communication channel in each morphology (e.g. median azimuth spread is 12° in Urban Macrocell (UMa) and 3° in Suburban Macrocell (SMa)). Such a model is problematic as it models a 2-way monostatic scattering channel (TRP-clutter-TRP) as a 1-way channel (TRP-RP), not justified by physics. The model results in narrow angle spreads, entirely at odds with expectation and observations: the environment is generally statistically homogeneous in azimuth, with no reason to expect strong return from a “special” direction towards an RP. It also predicts backscatter power to increase with increasing antenna gains at both Tx and Rx, at least for antenna beamwidth wider than channel angle spread. That is at odds with indoor backscatter observations [18] as well as theoretical predictions using a radar equation, integrated over clutter distribution, also in [18]. The model in [9] prescribes additional, dense cluster components, attenuated by 25 dB relative to RP contributions, thus producing little impact. At present, the standard model in [9] does not contain temporal variation of clutter.

In [18] a statistical model for monostatic (or quasi-monostatic) clutter backscatter based on indoor measurements was presented. Model components included room-size dependent average backscatter power and lognormally distributed azimuthal variation. The measured indoor environment was essentially static. Here we examine an urban street canyon environment which exhibits temporal variations.

In this work we propose a measurement-based model for monostatic clutter backscatter in dynamic outdoor urban environment. Narrowband measurements of scattered power as a function of azimuth and time at 140 GHz were collected in two environments. First, sensing antennas were placed 8 m above a busy Manhattan street at 5 locations, emulating a lamppost mounted “microcellular” base station used as radar. A second data set, labeled “street level”, had the sensing antennas placed on a cart 1 m above the street, emulating a traffic sensing arrangement. Key quantity of interest is the backscatter power ratio, defined as the ratio between the receive backscattered power and transmit power, with antenna gains removed as appropriate. Collected data was analyzed to formulate a model representing observed azimuthal and temporal variation of backscattered power ratio.

Based on the data, a statistical model of the backscatter power ratio is represented as varying over azimuth about an average, with variation statistics derived from measurements. Key results include:

- Average measured clutter backscatter power ratio was found to be well modeled (under 3.3 dB RMS error) by a simple, previously derived formula, dependent on distance to major clutter (building wall).
- Variation of clutter backscatter with azimuth about its local average was found to be well represented as having a lognormally distributed amplitude. The resulting cumulative distribution function (CDF) of the variation statistics is within only 1 dB of the observed statistics.
- Temporal fluctuation for various antenna aims and locations were found to be well modeled by a Rician distribution, with lognormally distributed K-factor and fluctuating part of the channel with coherence time below 0.6 sec.
- For any direction, deviation from mean power and temporal K-factor were found to be correlated with a 0.47-0.73 correlation coefficient, indicating that directions of strong backscatter return also tended to have less power fluctuation. For any particular direction, street-level backscatter exhibited higher variability with a lower K-factor and lower correlation between the K-factor and deviation from local mean.
- The dynamic statistical model does not require a detailed environmental layout, aiming to reproduce backscatter clutter statistics, as opposed to a deterministic response.

## II. MEASUREMENT EQUIPMENT

The sensing measurement set-up was adopted from a narrowband channel sounder [14]. The resulting sensing sounder consisted of a transmitter and receiver placed on the same cart, separated vertically, isolated from each other by absorbing foam to avoid direct signal leakage. The measurement system is meant to emulate a backscatter-sensing micro base station placed on a lamppost or a building wall, in the street canyon below building heights. A 140 GHz CW tone at 22 dBm was transmitted from an omnidirectional antenna, with gain $G_T$=2 (3 dBi), having a vertical beamwidth of about 60°. This transmitter was supported by hollow plastic cylinder placed on top of a plastic cart (Fig. 1). The plastic bin was determined through a calibration measurement to have under 1 dB of loss. The receive antenna was a 30 dBi horn with 2° half-power beamwidth performing a full azimuthal rotation every 0.6 seconds. Received power is measured and recorded at 4000 samples/sec. The receiver has 70 dB dynamic range. The spinning horn receive antenna was placed on the cart (inside the plastic bin), 0.9 m below the omni transmitter, (Fig. 1).

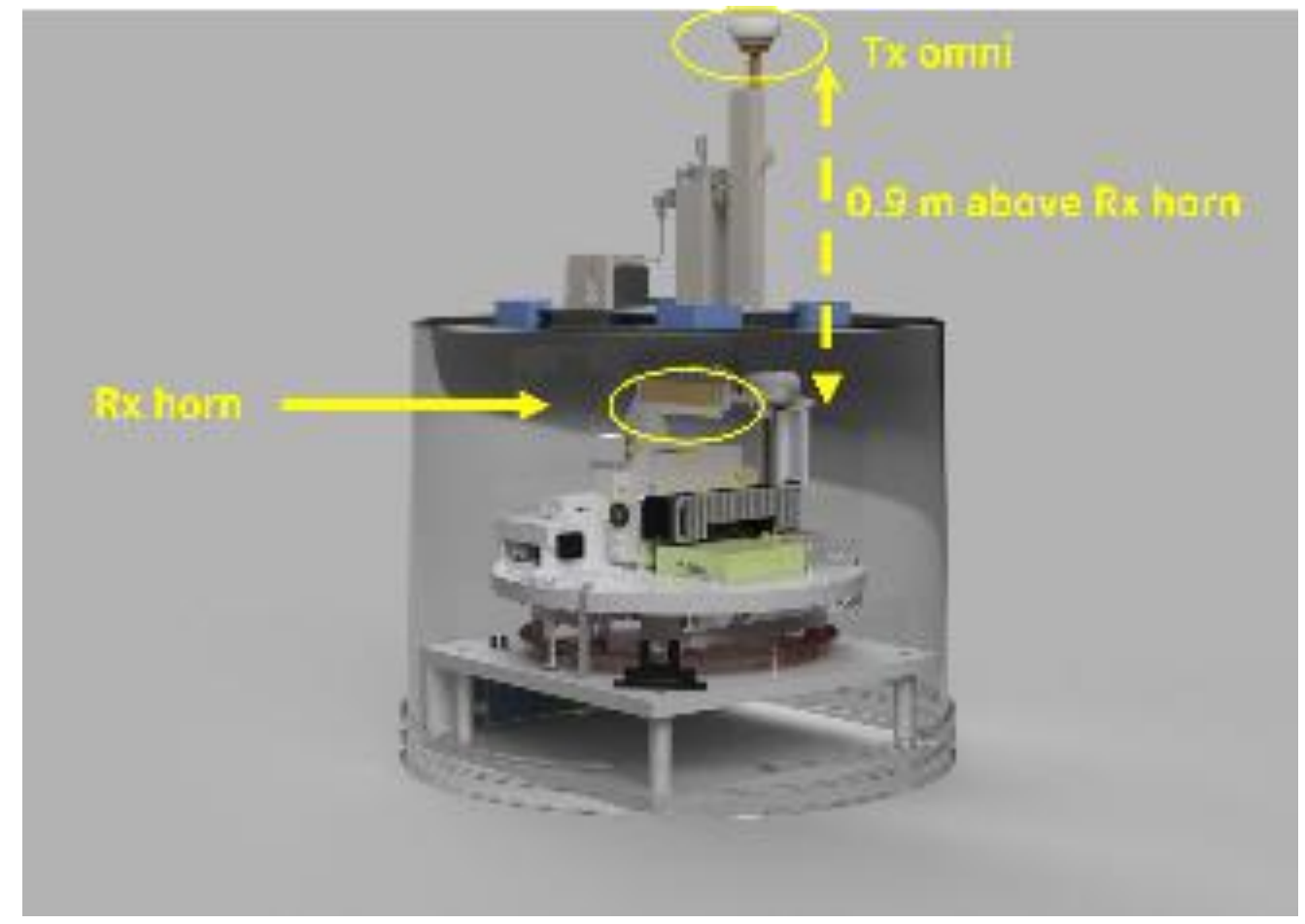

Fig. 1. Quasi-monostatic antenna arrangement, with omnidirectional transmit antenna placed about 1 m above a spinning horn receiver.

## III. MEASUREMENT ENVIRONMENT

Measurements were collected at 5 “microcellular” radar locations, shown in Fig. 2a, with the sounder cart placed on an overpass 8m above street level in Manhattan, emulating a lamppost-height microcellular sensing base station. At each location, 150 fluctuating power records were collected, measured every 1 deg in $-75 < \phi < 75$ azimuth, for a total of 750 time records. Each time series contains power values measured over 3-10 minutes, every 0.6 sec, for a total of 488234 power values.

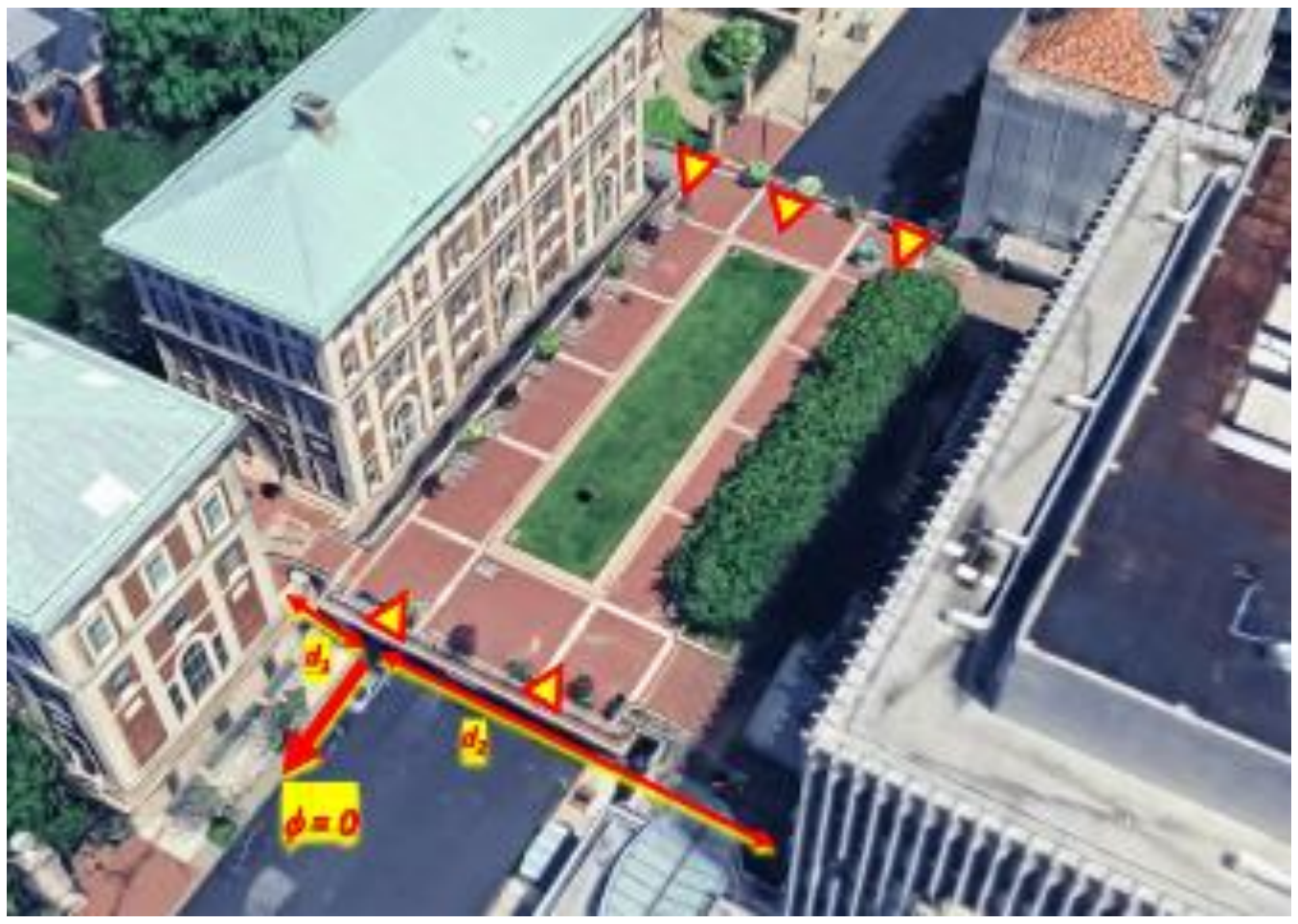

Fig. 2a. Five "microcellular" radar locations 8 m above Amsterdam Ave in New York.

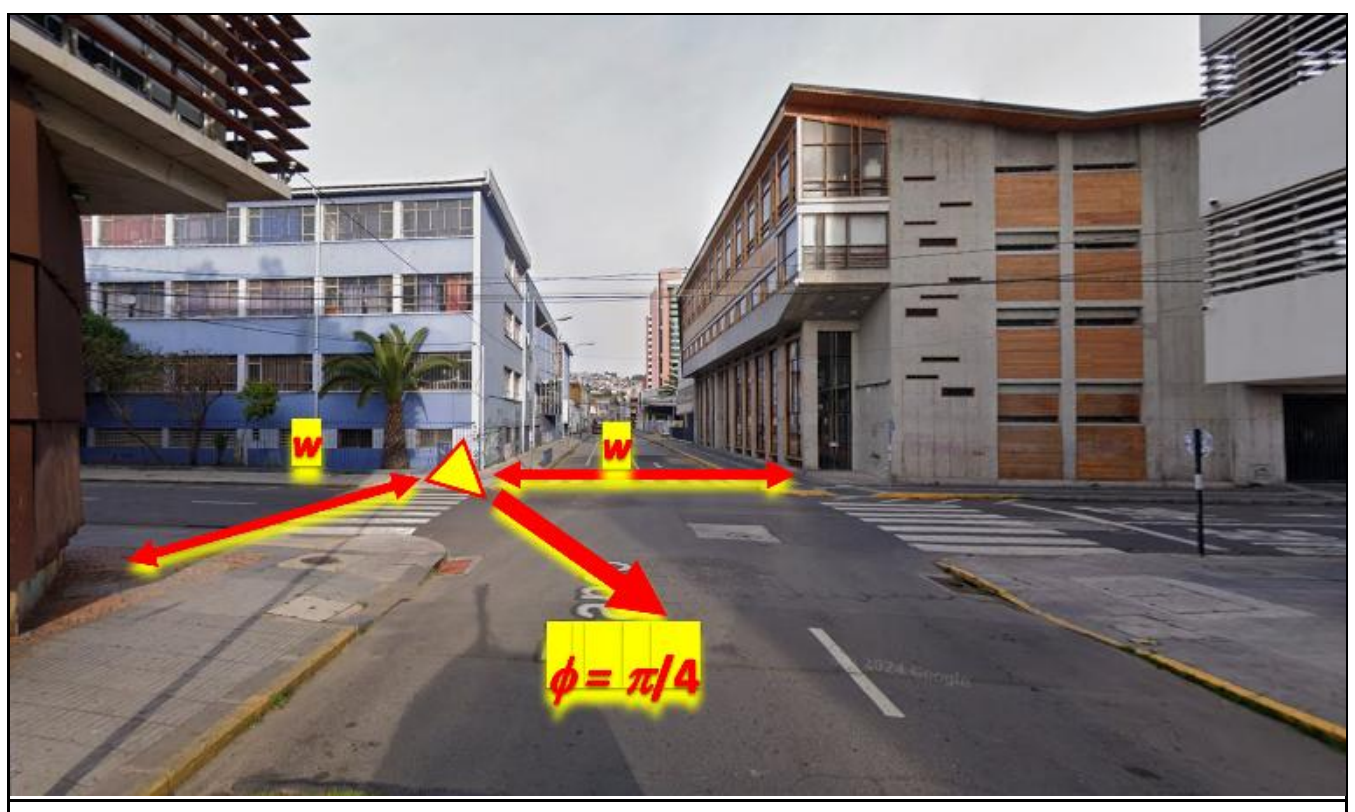

Fig. 2b. Street-level backscatter geometry 1m above ground at an urban intersection in Valparaiso, Chile.

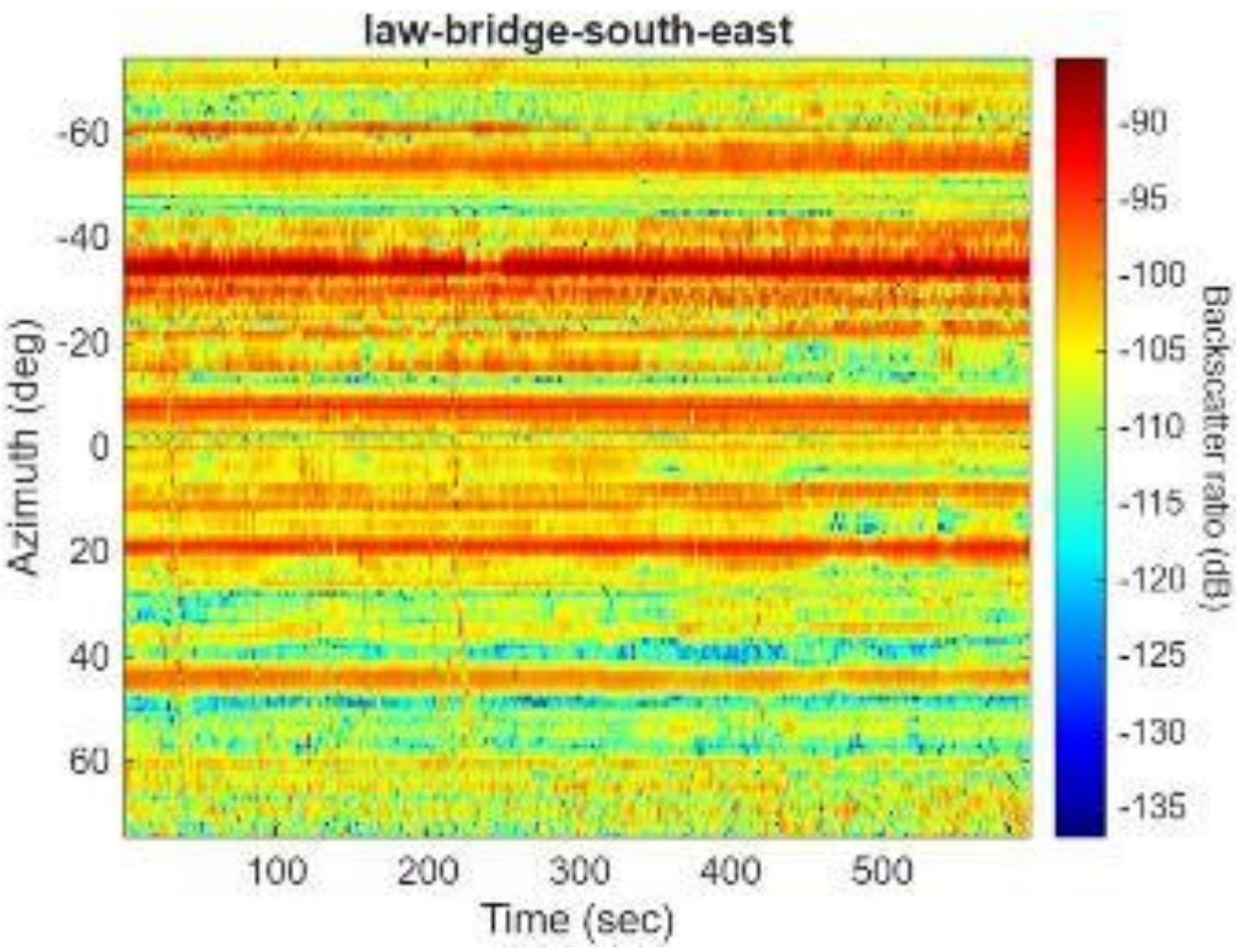

Fig. 3. Sample measured backscattered power ratio vs. azimuth and time.

A second data set, labeled "street level", collected similar data using a cart 1 m above the street, at 12 locations facing 3 urban intersections in Valparaiso, Chile, emulating a traffic-sensing radar. One such placement is illustrated in Fig. 2b.

In both cases, visually observed clutter consisted of building walls (generally with uneven façades), as well as parked and moving vehicles, pedestrians and street furniture.

Backscatter power as a function of time and azimuth collected at one of the microcellular locations (lower right of Fig. 2a) is shown in Fig. 3.

## IV. Average Backscattered Power

Measured time-averaged azimuthal pattern $\left\langle P_{\text{back}}\left(\phi,t\right)\right\rangle_t$ at a particular location is illustrated in Fig. 4, after averaging time-azimuth backscatter shown in Fig. 3. As the presumed deployment would be to use a wall mounted antenna to sense the intersection environment in front of it, all the processing and display is only done for the 150° field of view, $-75^o < \phi < 75^o$, where azimuth $\phi$ = 0 is the direction is as indicated in Fig. 2. This excludes scattering from objects immediately behind radar as they would not be illuminated by a wall mounted antenna. Average backscatter power ratio, with average performed over both time and azimuth:

$$\left\langle P_{\text{back}}\right\rangle_{\phi,t} = \frac{\left\langle P_{\text{rec}}\left(\phi,t\right)\right\rangle_{\phi,t}}{P_{\text{T}} G_T} \tag{1}$$

is indicated as a dashed semicircle in Fig. 4. Average backscattered power ratio (1) for clutter was shown [18] through a simple theoretical derivation and confirmed through indoor measurements, to decrease with increasing room size and to be independent of the antenna pattern of the (high gain) receive antenna. Specifically, it was found in [18] that average backscatter power ratio from an extended scatterer (such as a rough wall or a collection of objects) at distance $d_{\text{s}}$ from the monostatic radar can be modeled as proportional to the average over distances to the walls bounding the sensing area:

$$\left\langle P_{\text{back}}\right\rangle_{\phi,t} = \left\langle \left|\Gamma_{\text{clut}}\right|^2 \left(\frac{\lambda}{4\pi d\left(\phi\right)}\right)^2 \right\rangle_{\phi} \tag{2}$$

where $\left|\Gamma_{\text{clut}}\right|^2$ is the power reflection coefficient dependent on material and geometric (e.g. roughness) properties of surrounding environment. For complex media it is often treated as a heuristic quantity, chosen as a fit to data [13]. Here, as indoors in [18], $\left|\Gamma_{\text{clut}}\right|^2 = 1$ was found to reproduce observed backscatter values accurately (under 3.3 dB rms error in this work). It is therefore omitted in equations that follow. Notably the average backscattered power ratio (2) is independent of the antenna pattern of the higher gain antenna, which is the receive antenna gain $G_{\text{R}}$ in the case of measurements conducted here. This makes (2) independent of the measurement apparatus, including antennas, and characteristic of the environment under study. Average backscattered power (2), as indicated, depends on the distribution of scatterers around the radar station, averaged over azimuth. In the case of a street with buildings bounding the sensing area on either side (Fig. 2a)

$$d(\phi) = d_s / \cos\phi, \quad d_s = \begin{cases} d_1, & |\phi| < \pi/2 \\ d_2, & |\phi| > \pi/2 \end{cases} \tag{3}$$

where $d_1, d_2$ are the shortest distances from the radar to left/right walls, respectively, illustrated in Fig. 5.

Using (3) to evaluate the average backscatter power ratio over azimuth angles $\phi$, leads to

$$\langle P_{\text{back}} \rangle_{\phi,t} = 0.25\left(\frac{\lambda}{4\pi d_1}\right)^2 + 0.25\left(\frac{\lambda}{4\pi d_2}\right)^2, \tag{4}$$

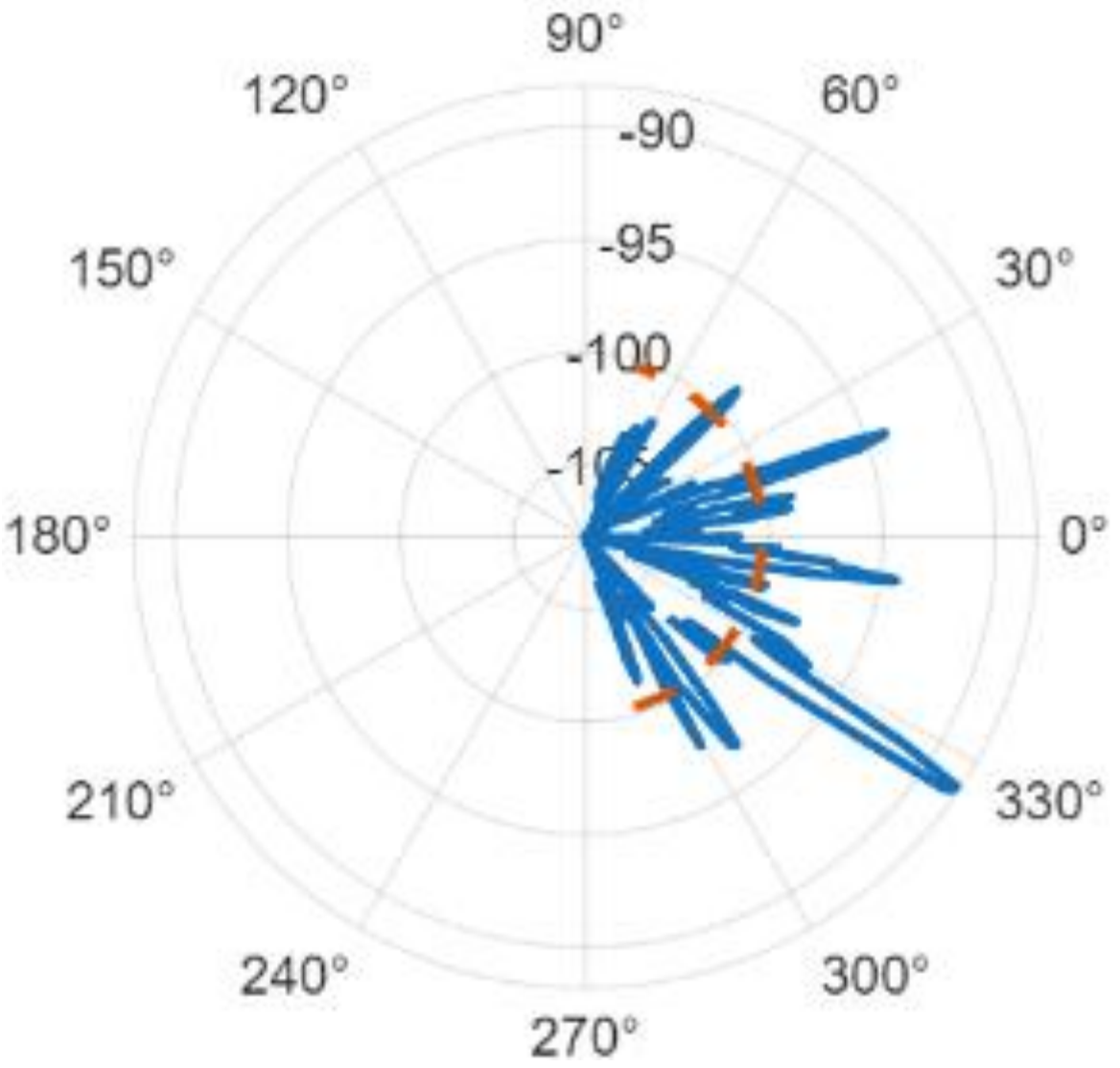

Fig. 4. Sample measured time-averaged backscattered power ratio vs. azimuth angle. Dashed line is average backscattered power ratio (1)

The width of the measured street was 25m, with radar distances (Fig. 5) from walls varying between $d_1 = d_2 = 12.5$ m for the location above the street median and $d_1 = 7$ m, $d_2 = 18$ m or $d_1 = 18$ m, $d_2 = 7$ m for locations closer to either side of the street (Fig. 2a).

Measured average backscatter ratio (1) at 5 midstreet locations (8m height) is plotted in Fig. 6a as solid circles. , The vertical bars around it indicate RMS variation (dB) as the receiver antenna scans the azimuth, illustrated in Fig. 4. The theoretical predictions (4) is indicated as open squares. The RMS error of theoretical prediction is found to be 2.0 dB. In [18] a similar formula was found to be similarly accurate at predicting average backscatter power across rooms of different sizes.

For "street-level" antennas facing the intersection in Fig. 2b, the distance to different points on the wall across the intersection (Fig. 5b) is

$$d(\phi) = w / \cos\phi \tag{5}$$

Where $w$ is the street width, $w$=9m in this case. To avoid scattering from the building immediately behind the cart (not of interest for traffic monitoring at the intersection), backscatter data at the intersection was limited to a 90° sector centered around the direction towards the middle of the intersection, corresponding to $\phi \in [\pi/4, \pi/2]$ and $\phi' \in [\pi/4, \pi/2]$, Fig. 5b. Evaluating the averaged (over azimuth) backscattered power ratio (2) using (5):

$$\begin{aligned} \langle P_{\text{back}} \rangle_{\phi,t} &= \left(\frac{\lambda}{4\pi w}\right)^2 \left( \int_{\pi/4}^{\pi/2} \cos^2\phi \, d\phi + \int_{\pi/4}^{\pi/2} \cos^2\phi' \, d\phi' \right) \Big/ (\pi/2) \\ &= (0.5 - 1/\pi)\left(\frac{\lambda}{4\pi w}\right)^2 \end{aligned} \tag{6}$$

Measured backscatter power, averaged over azimuth and time at eighteen 10-minute records, was collected at 7 distinct locations, 1 m above street level at 3 intersections in Valparaiso, Chile. It is seen in Fig. 6b that observed average backscatter power is represented by (6) with 3.3 dB RMSE.

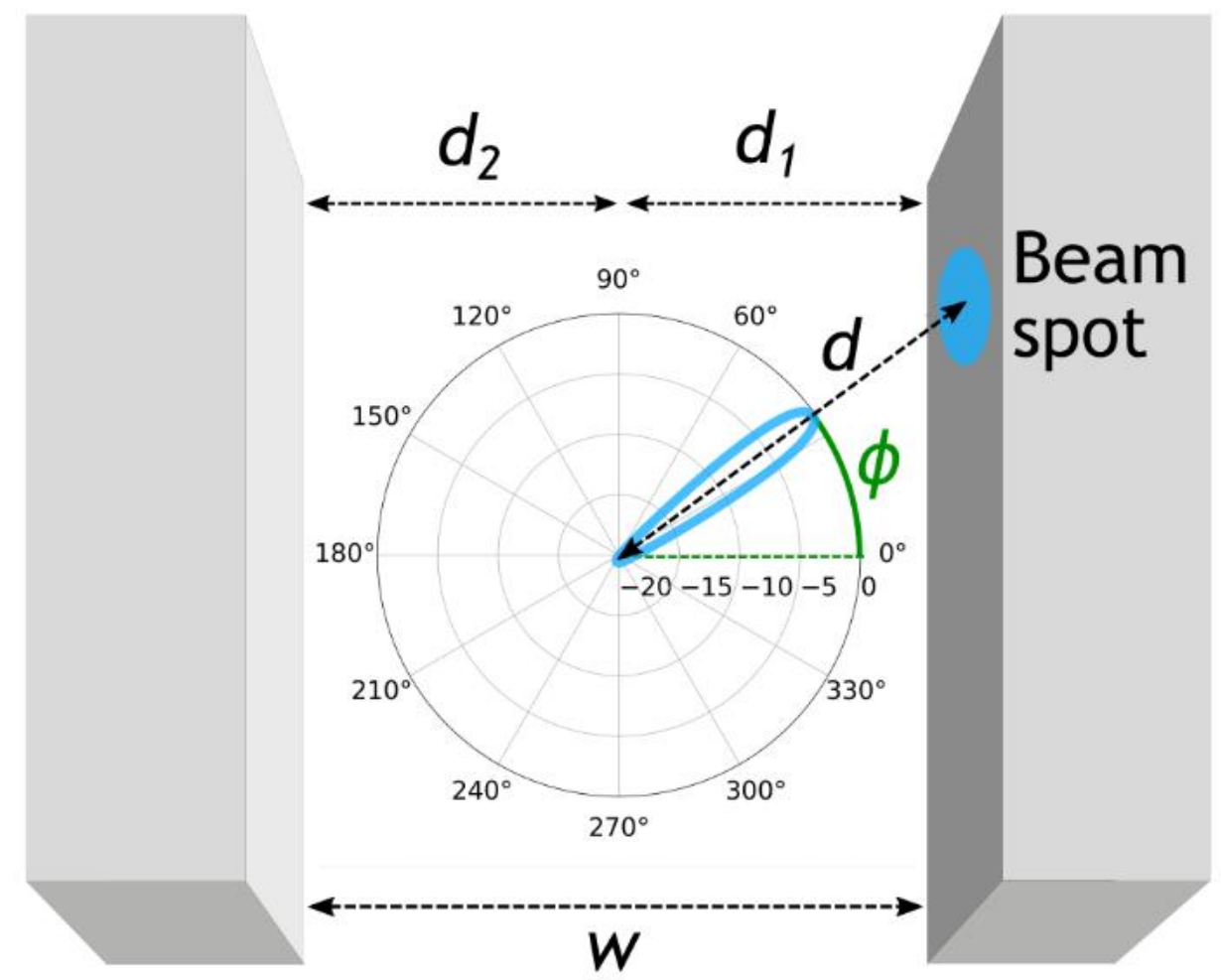

Fig. 5. Typical backscatter measurement geometry in midstreet deployments (Fig. 2a), for rough building walls along the street. Main parameters: distances $d_1$ and $d_2$ from radar to buildings on either side of street.

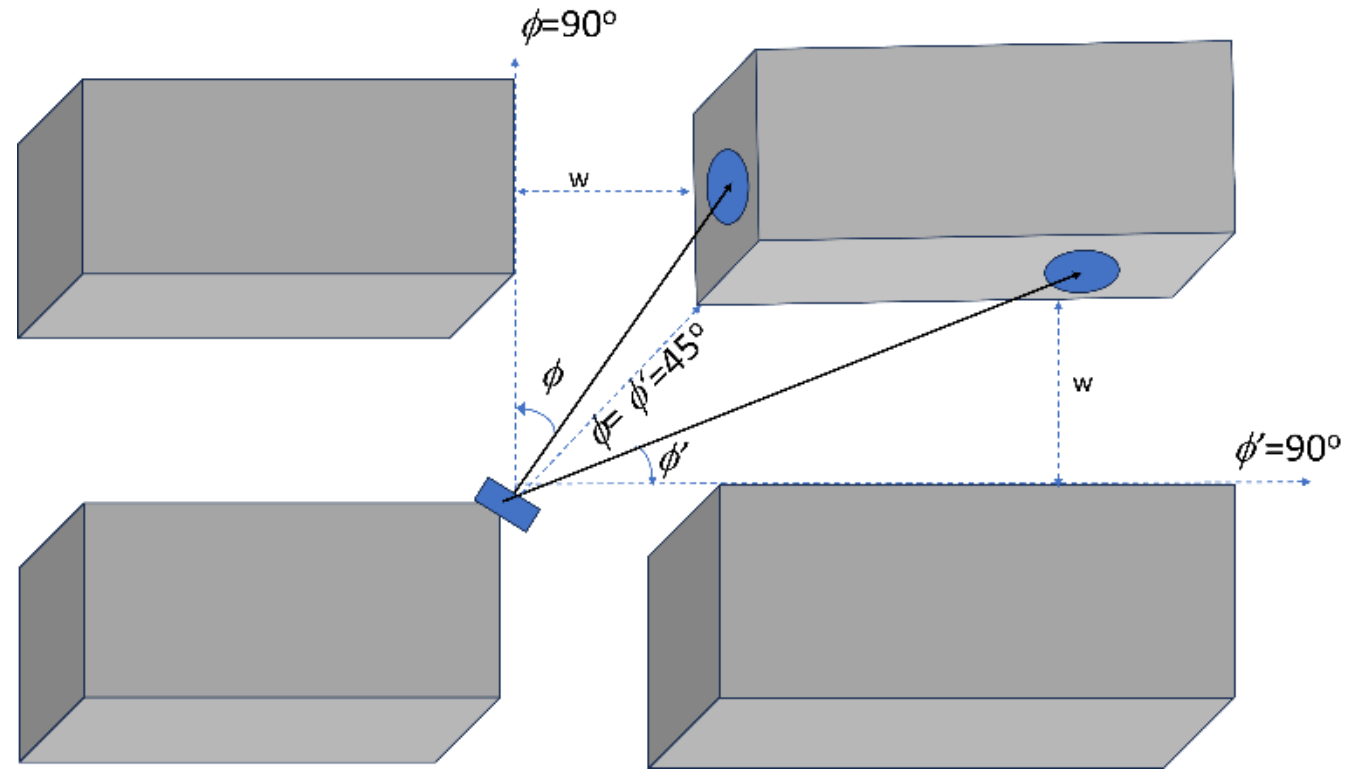

Fig. 5b. Backscatter measurement geometry at an intersection (Fig. 2b), for rough building walls along the street

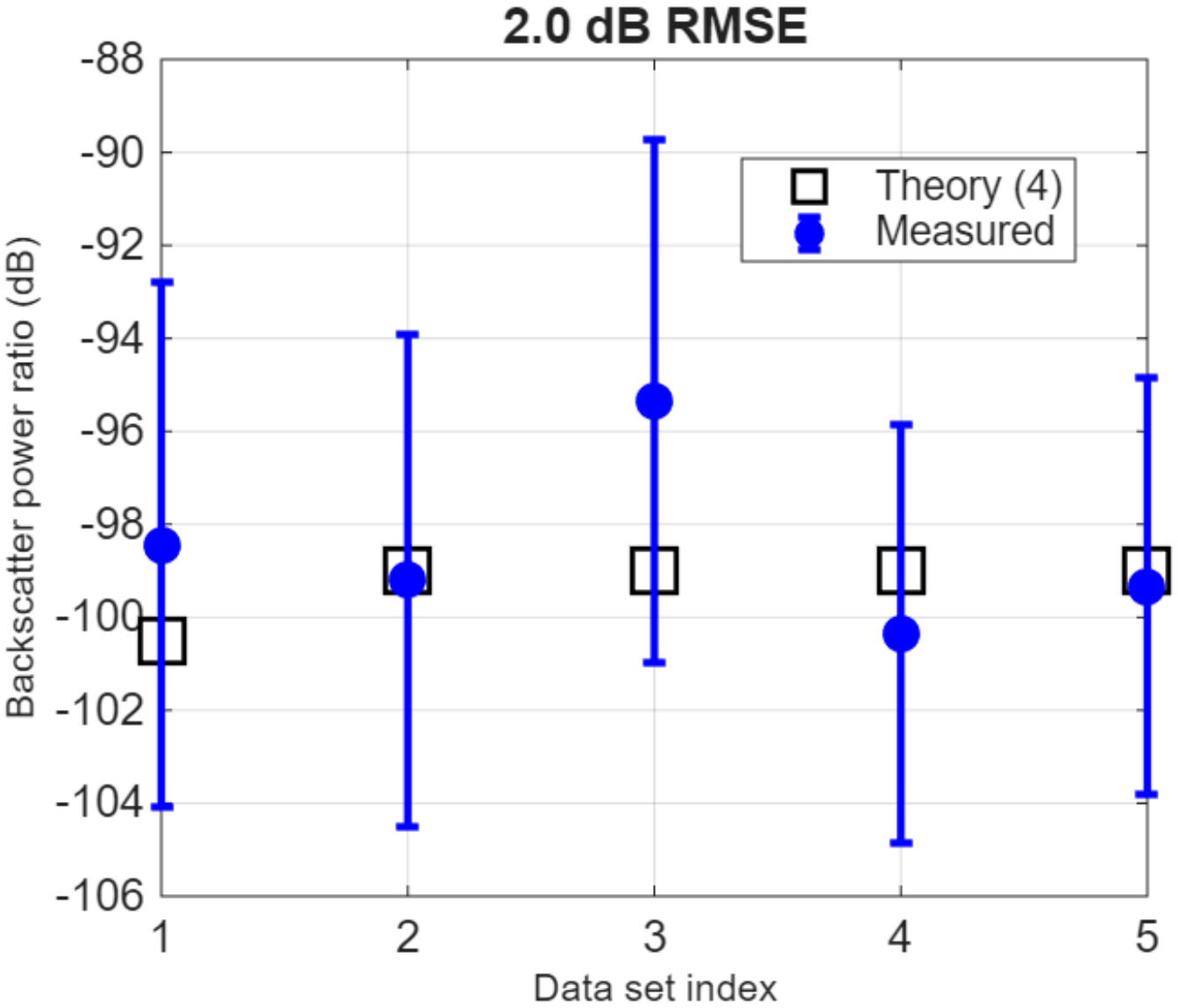

Fig. 6a. Averaged backscattered power ratio (filled circle) measured (1)at microcellular locations. Predictions (4) (open squares). Vertical bars indicate rms variability with azimuth.

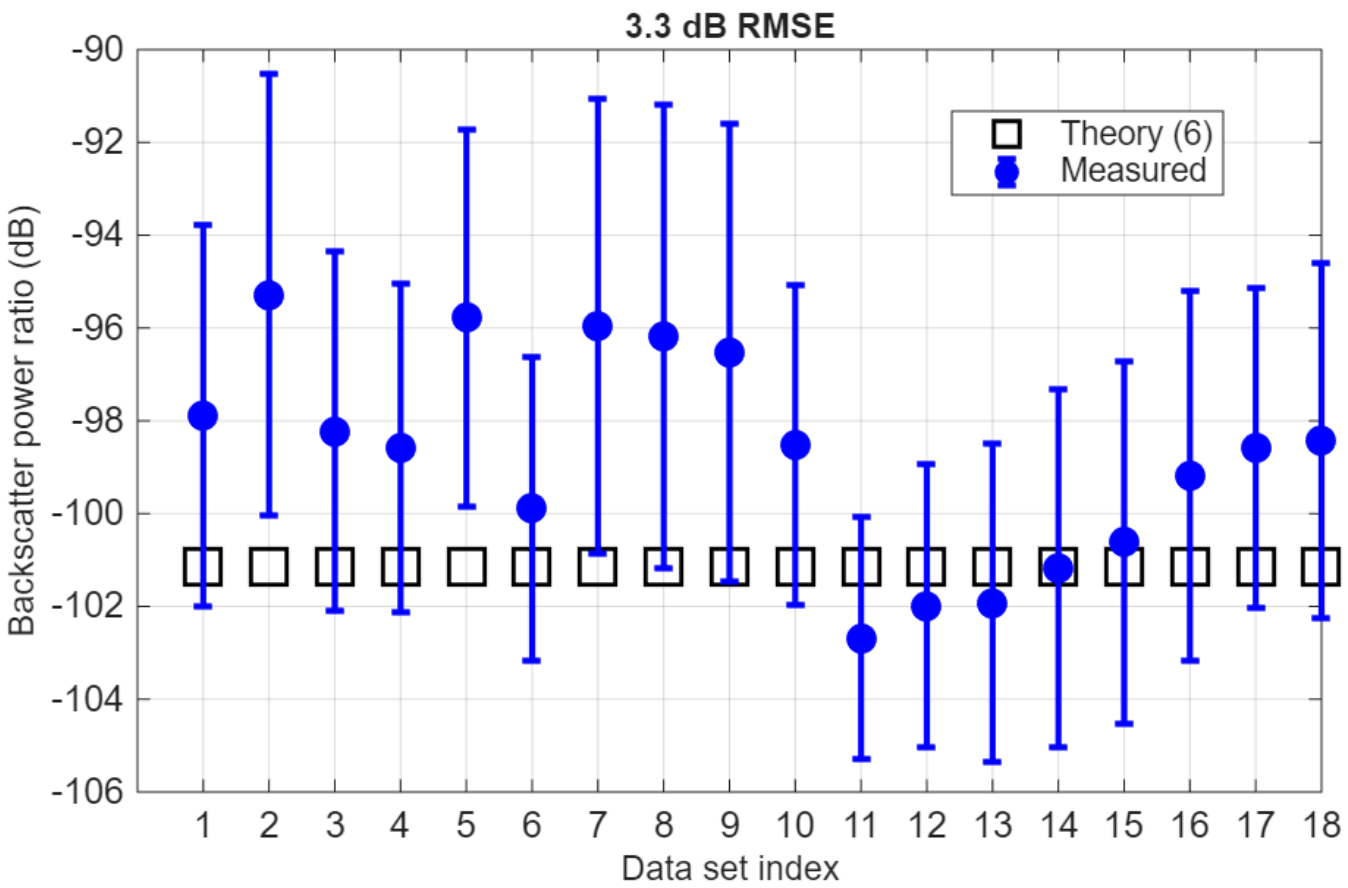

Fig. 6b. Measured averaged street-level backscattered power ratio (1) (filled circle). Open squares are predictions using (6). Vertical bars indicate rms variability with azimuth

## V. OBSERVED AZIMUTH VARIATION STATISTICS

As illustrated in Fig. 4, measured backscattered power exhibits variation with azimuth around average power. The distribution of observed power variations relative to observed average power (1):

$$P_{\text{rel}}(\phi) = \frac{\langle P_{\text{back}}(\phi,t)\rangle_t}{\langle P_{\text{back}}\rangle_{\phi,t}} \tag{7}$$

is plotted as a solid line in Fig. 7 for the “lamppost” data. The empirical standard deviation of azimuthal variation (7) (expressed in dB) for microcellular data is found to be 5.3 dB. It is seen in Fig. 7 to be within 1 dB of the lognormal distribution (dashed line) $10\log_{10} P \sim N(\mu,\sigma)$ ,with $\sigma = 5.3$ dB and mean $\mu = -10\log_{10} e^{(0.1\sigma\ln 10)^2/2}$ , dependent on σ to assure the mean of $P$ (simulated relative power) is 1, just as measured $\langle P_{\text{rel}}(\phi)\rangle_\phi = 1$, following from (1) and (7). Similar observations are obtained for street-level data, with $\sigma = 4.0$ dB .

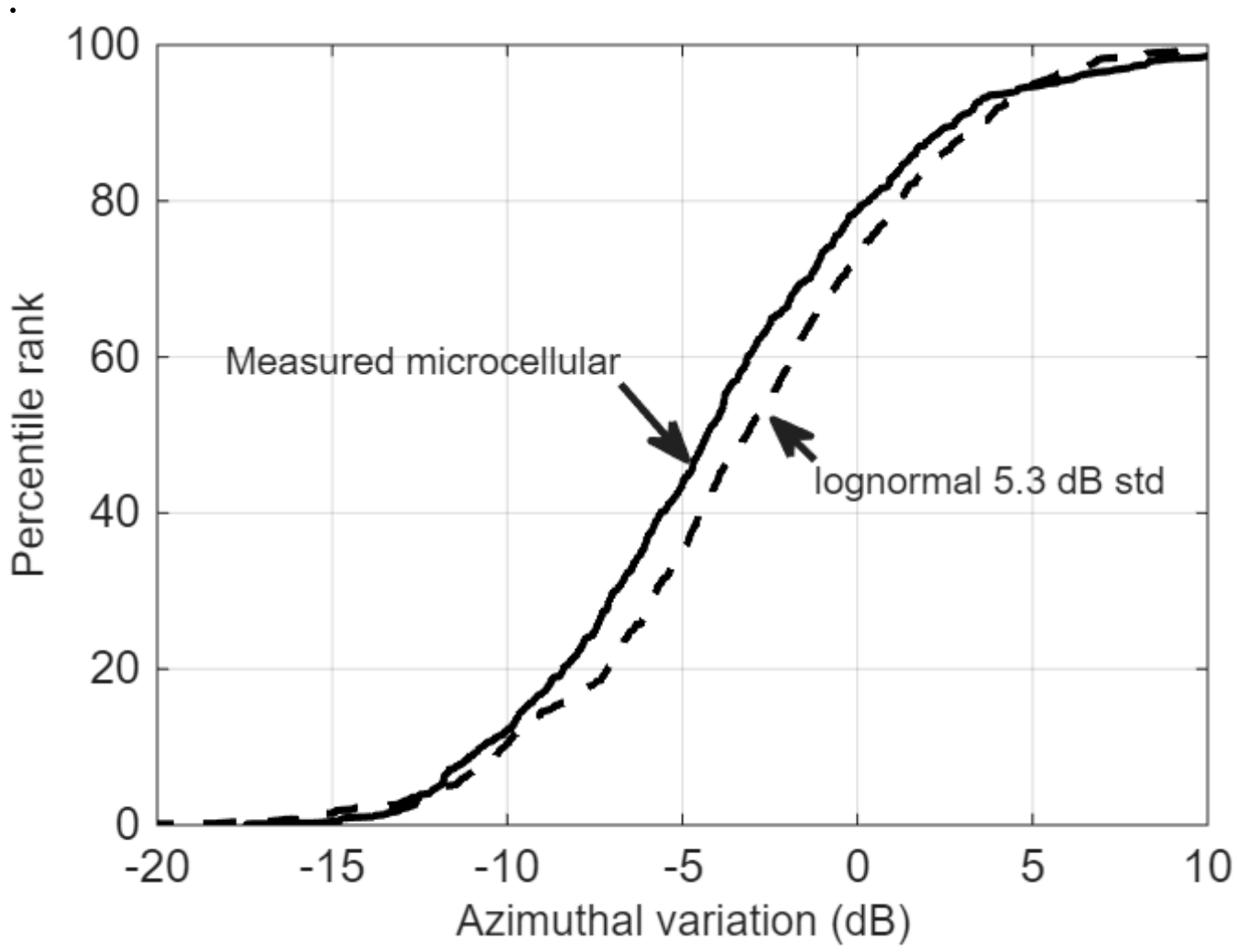

Fig. 7. Cumulative distribution of variation with azimuth of measured backscattered power ratio (7) relative to mean power (over time and azimuth) at microcellular locations. Dashed line is the lognormal distribution with same RMS variation of 5.3 dB as the observation.

## VI. OBSERVED TEMPORAL VARIATIONS

Temporal power fluctuation is defined here as variation of power as a function of time for each antenna aim direction relative to time-averaged receive power at each location:

$$P_{\text{fluct}}(\phi,t) = \frac{P_{\text{rec}}(\phi,t)}{\langle P_{\text{rec}}(\phi,t)\rangle_t} \tag{8}$$

An illustration of the observed temporal fluctuation (8) computed from single 600 sec data record in Fig. 3, is shown in Fig. 8 vs. azimuth $\phi$ and time $t$. Strong up-fades (>10 dB, red) and down-fades (>15 dB, blue) appear fairly rare and widely distributed (not very clustered) across azimuth and time.

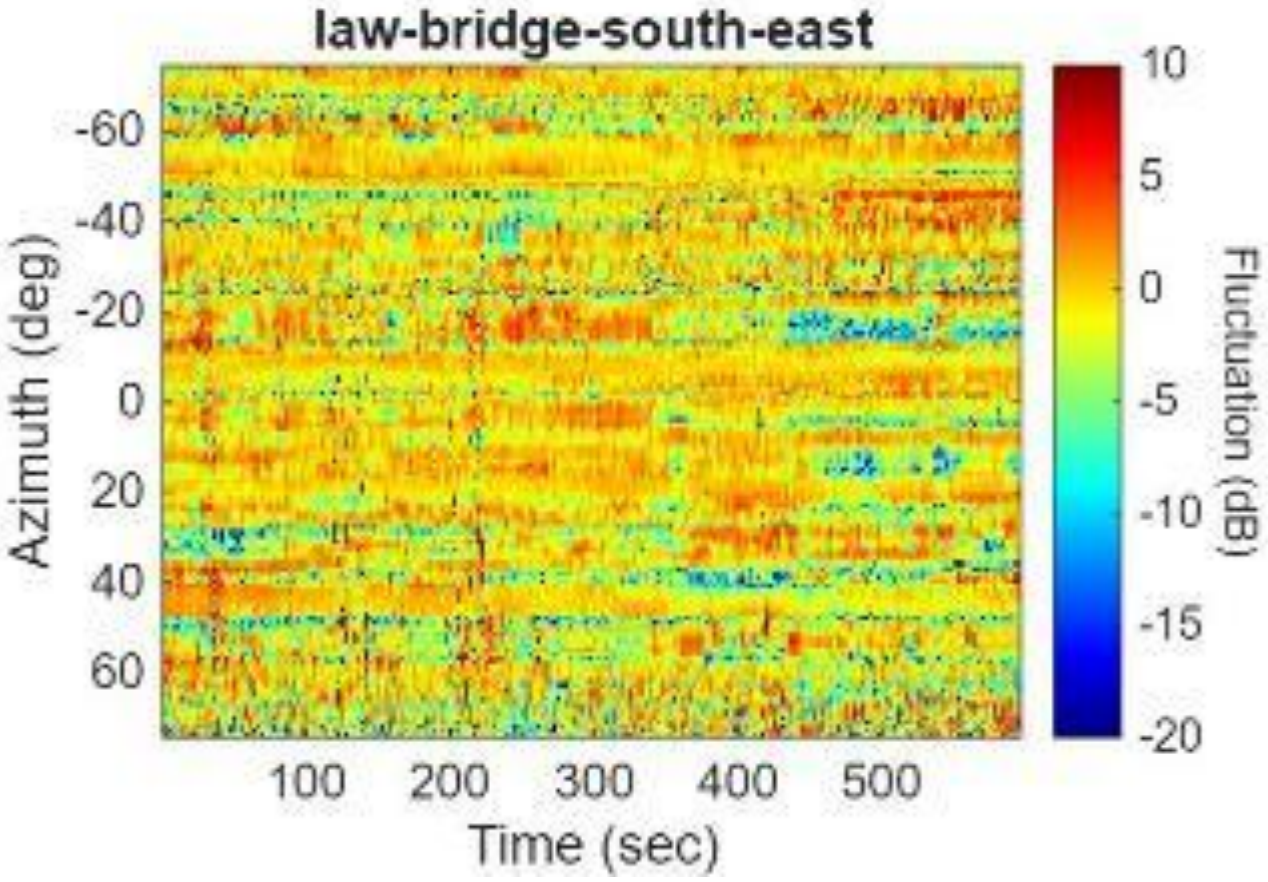

Fig. 8. Backscattered power fluctuation (dB) relative to mean over time (7) in each azimuth direction, measured on southeast corner of the overpass.

More specifically, backscatter fluctuation around mean, measured when looking south along Amsterdam Ave ($\phi = 0$) from southeast end of the overpass is shown in Fig. 9. The bottom plot shows measured backscattered fluctuation as a function of time, while its empirical distribution in upper plot is seen to be well approximated by a Rician distribution with a K-factor of 13.9 dB, estimated using the moment-method [19].

Here K-factor is interpreted as a ratio of constant power scattered from stationary objects, like buildings and parked cars to time-fluctuating power, scattered from moving objects, such as moving vehicles, pedestrians and foliage driven by wind. Distribution of each power time series was compared to corresponding Rician distribution, as in the example in Fig. 9. It was found that 80% of time records were within 1.3 dB of corresponding Rician distribution.

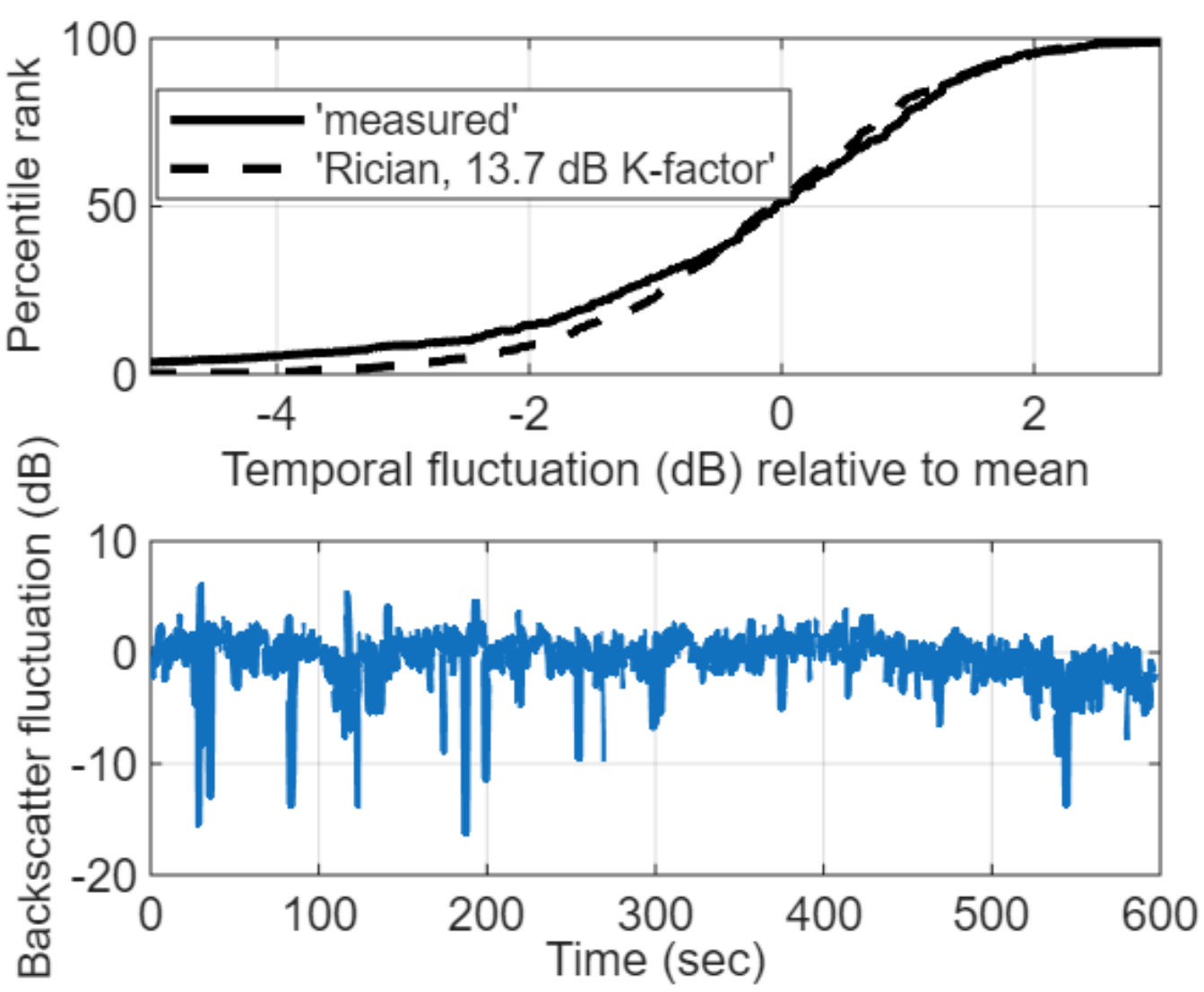

Fig. 9. One example of backscattered power fluctuation (dB) relative to mean, looking south along Amsterdam Ave. ($\phi = 0$) at microcellular location. Distribution at top, time series at bottom.

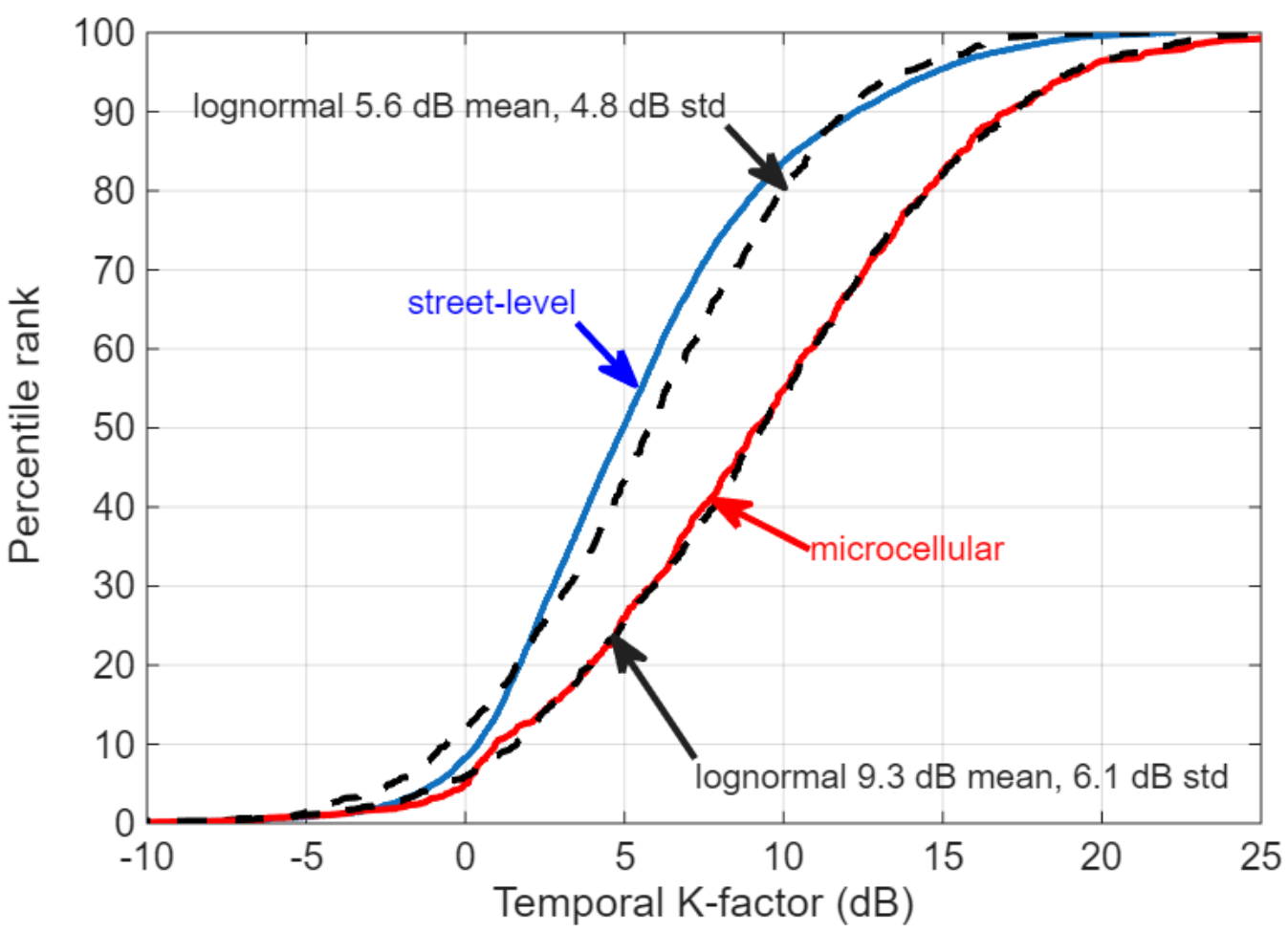

Fig. 10. Cumulative distribution of temporal K-factors.

Distribution of Rician K-factors extracted from 750 time series at 8m height (mimicking microcellular scenario), is plotted in Fig. 10 and was found to be within 0.5 dB of a lognormal distribution, $K \sim N(9.3, 6.1)$ (dB). Also shown is the distribution of 1m height street-level K-factors, extracted from 9000 time-series measured in different directions and at different locations. The street-level data K-factor is seen to be well represented by a lognormal distribution $N(5.6, 4.8)$ (dB). Lognormal distribution of K-factor was found to be representative of fluctuation of communication channels in [20].

Coherence time of backscatter power fluctuation relative to local mean is assessed by examining a time autocorrelation of relative power fluctuation. For the power time series shown at the bottom of Fig. 9, the autocorrelation coefficient is

$$C(t) = \frac{\left\langle \left( p_{\text{rec}}(t_1) - \langle p_{\text{rec}}(t_1) \rangle_{t_1} \right) \left( p_{\text{rec}}(t_1 + t) - \langle p_{\text{rec}}(t_1) \rangle_{t_1} \right) \right\rangle_{t_1}}{\sqrt{\left\langle \left( p_{\text{rec}}(t_1) - \langle p_{\text{rec}}(t_1) \rangle_{t_1} \right) \right\rangle_{t_1} \left\langle \left( p_{\text{rec}}(t_1 + t) - \langle p_{\text{rec}}(t_1) \rangle_{t_1} \right) \right\rangle_{t_1}}} \tag{9}$$

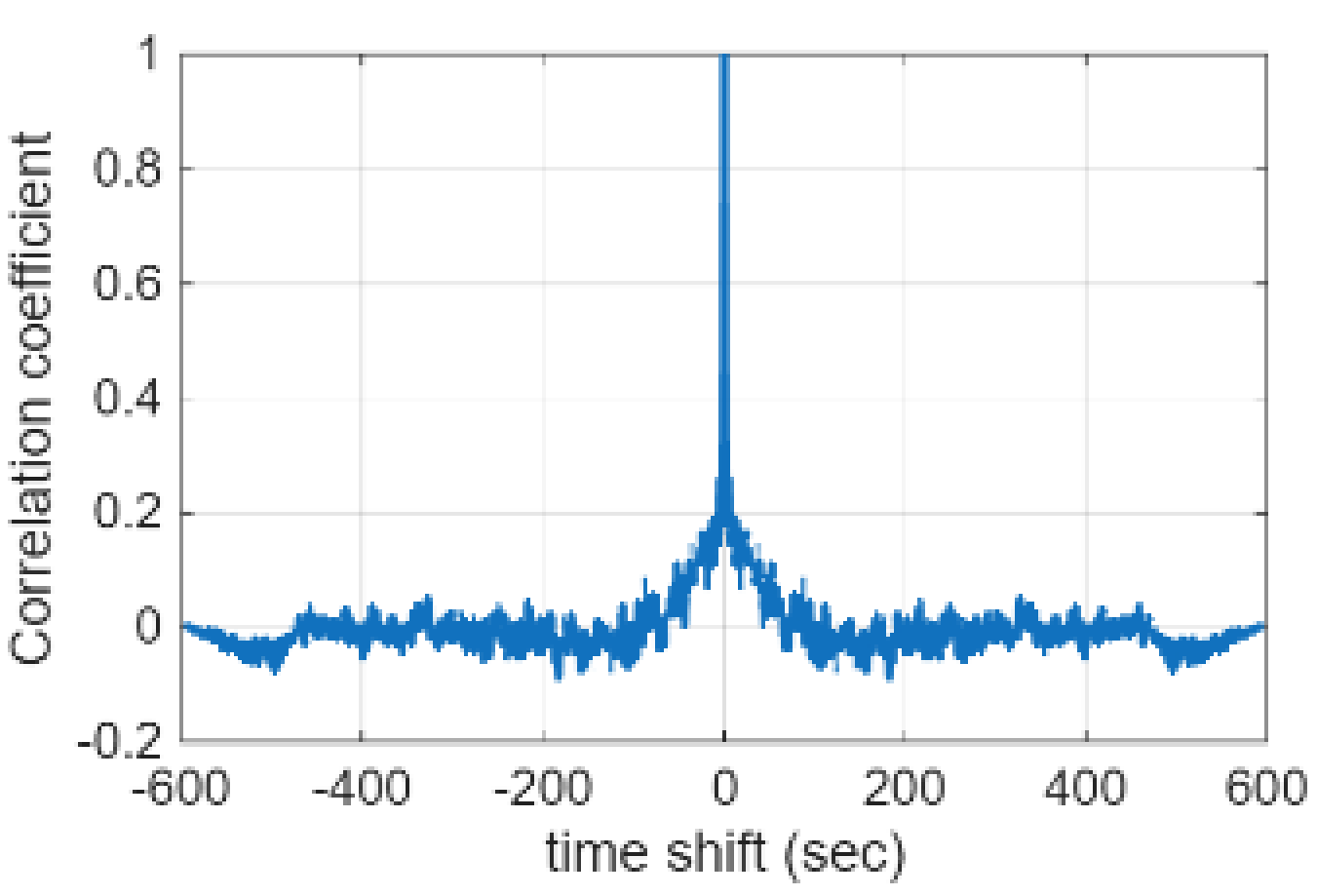

Fig. 11. Autocorrelation coefficient of backscatter power fluctuation time series at bottom of Fig. 9.

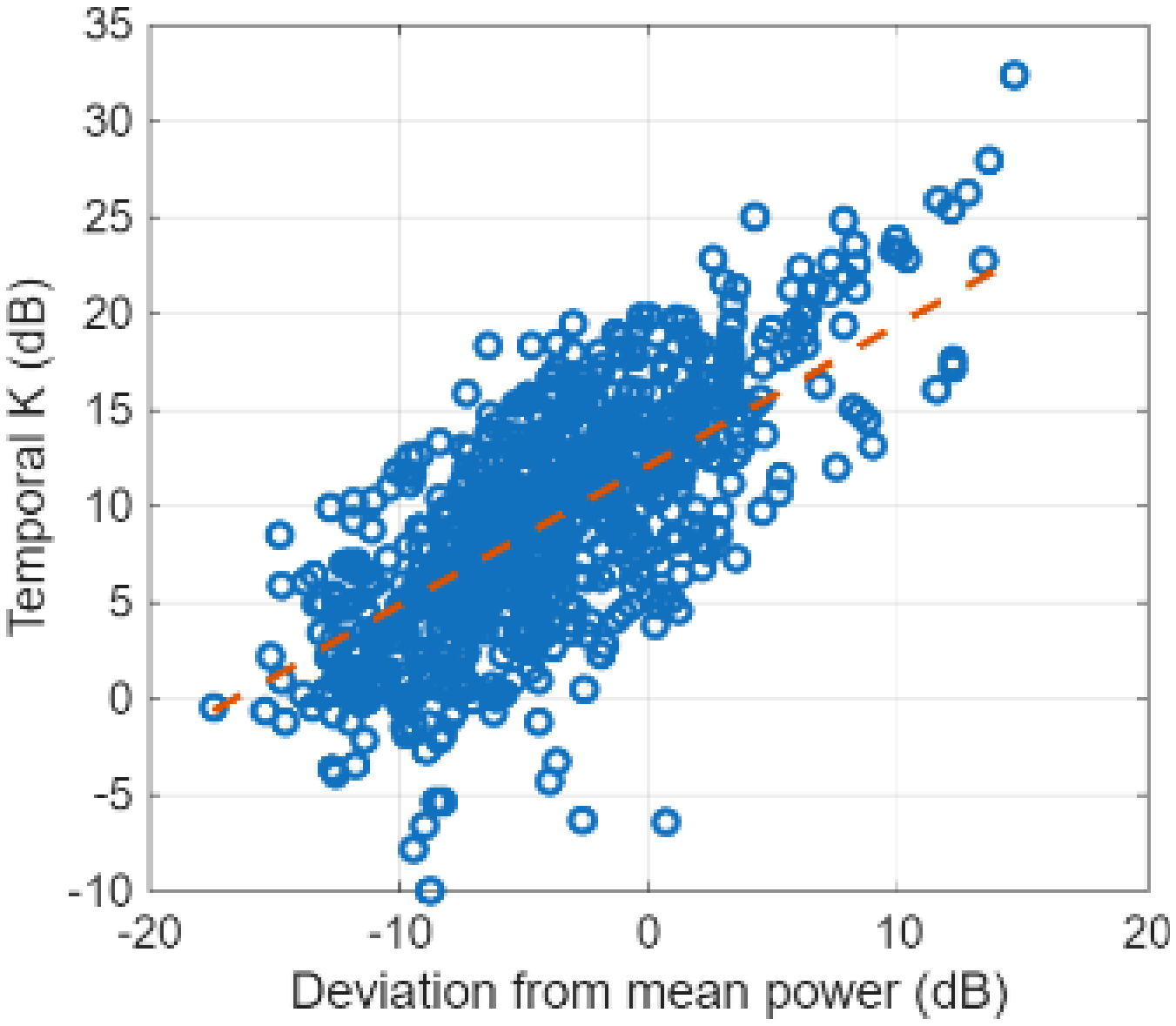

Fig. 12. Observed deviation from mean power (7) and temporal K-factor across all measured directions and all 5 locations at microcell height. Dashed line is the trend line with 0.73 correlation coefficient.

The observed time autocorrelation coefficient of backscatter power fluctuation, plotted in Fig. 11, decorrelates very rapidly, dropping to correlation of 0.2 in 0.6 sec. Since the data collected at 100 revolutions per minute (RPM) in this study recorded power in any particular direction every 0.6 sec, the finding is that temporal power fluctuations are faster than 0.6 sec. Other locations were found to be similarly rapidly varying.

We now investigate the relationship between relative backscatter power deviation (7) from mean in a particular direction and temporal K-factor in that direction. Observed deviations from mean (azimuth fade) and temporal K-factor in all azimuth directions and at all 5 microcellular locations are plotted in Fig. 12. These two quantities are found to have a correlation coefficient of 0.73. This suggests that non-fluctuating echoes are also strong, coming from large, stationary objects, like buildings, while moving objects, like cars, produce weaker, time-fluctuating echoes. Similar analysis for street-level data showed a correlation of 0.47.

## VII. Dynamic statistical backscatter model

The above findings are now used to formulate a statistical backscatter model to allow generation of synthetic backscatter as a function of azimuth and time at outdoor intersections. In [18] a model for indoor backscatter as a function of delay and azimuth was formulated based on measurements of power vs. azimuth and power vs. delay behavior reported in literature. In the present work, the observed dynamic (i.e. time-varying) distribution of power vs. azimuth are now used to define a complex backscatter channel response $h_{\text{back}}(\phi,t)$ as a function of azimuth $\phi$ and time $t$:

$$h_{\text{back}}(\phi,t) = \sqrt{P_0(d_s)}\sqrt{10^{P(\phi)/10}}\, h_{\text{fluct}}(\phi,t) \tag{10}$$

where the average backscatter power is expressed as a function of street width $d_s$ by a representative formula

$$P_0(d_s) = 0.25\left(\frac{\lambda}{4\pi d_s}\right)^2 \tag{11}$$

Formulas appropriate for deployments off the street median and at intersections are given by (4) and (6), respectively. As described in the previous section, in a particular direction, the average power deviation from mean and temporal K-factor are correlated, lognormally distributed variables (i.e. directions from which backscattered power is particularly strong also tend to be steady). The corresponding model is:

$$\begin{pmatrix} P(\phi) \\ K_{\text{dB}} \end{pmatrix} = \begin{pmatrix} \mu_P \\ \mu_K \end{pmatrix} + \begin{pmatrix} \sigma_P & 0 \\ 0 & \sigma_K \end{pmatrix} \begin{pmatrix} 1 & \rho_{P-K} \\ \rho_{P-K} & 1 \end{pmatrix}^{1/2} \begin{pmatrix} \xi_P \\ \xi_K \end{pmatrix}, \tag{12}$$

where $\xi_P, \xi_K \sim CN(0,1)$ are iid random quantities. Means $\mu$, standard deviations $\sigma$ and correlation coefficients $\rho_{P-K}$ of observed power deviations and temporal K-factors are summarized in Table I.

Table I. Means $\mu$, std. deviations $\sigma$ and correlation coefficients $\rho_{P-K}$ of observed azimuthal power variations $P(\phi)$ and temporal K-factor $K_{\text{dB}}$

| Environment (height) | $\sigma_P$ | $\mu_P$ | $\sigma_K$ | $\mu_K$ | $\rho_{P-K}$ |
|---|---|---|---|---|---|
| Microcellular (8m) | 5.3 dB | -3.2 dB | 6.1 dB | 9.3 dB | 0.73 |
| Street-level (1m) | 4.0 dB | -1.9 dB | 4.8 dB | 5.6 dB | 0.47 |

Random temporal fluctuation $h_{\text{fluct}}(\phi,t)$ in (10) is modeled as a Rician distribution with parameter $K$

$$\begin{aligned} h_{\text{fluct}}(\phi,t) &\sim \sqrt{\frac{K}{K+1}}e^{i\varphi(\phi)} + \sqrt{\frac{1}{K+1}}\eta(\phi,t) \\ \varphi(\phi) &\sim U(0,2\pi), \quad \eta(\phi,t) \sim \mathbb{CN}(0,1) \end{aligned} \tag{13}$$

where $K = 10^{K_{\text{dB}}/10}$ using $K_{\text{dB}}$ generated in (12). Small scale randomness is generated via a random phase $\varphi(\phi)$, iid for every direction $\phi$ and complex Gaussian $\eta$ iid for different direction $\phi$ and time $t$.

The measurements collected using the antenna with $2^0$ azimuth HPBW, recorded at 100 RPM allowed measurement of backscattered power from a particular direction every 0.6 sec. The simplest way to include that in a simulation is to generate the channels as per formulas above using similar granularity, of 1° in azimuth and 0.6 sec in time. The model can be used to generate channels of similar or coarser granularity in azimuth and time.

The impact of using a directional antenna can be included by circularly convolving the channel response (10) with the field angular pattern of the antenna, as described in [18]. The resulting simulation would be valid only for antennas of same or wider beamwidth than the 2° antenna used to collect these measurements.

## VIII. Conclusions

A simple and accurate statistical monostatic radar channel model for outdoor clutter is developed for urban microcellular and street-level environments, based on measurements. A narrowband 140 GHz sounder was used in a quasi-monostatic antenna arrangement with an omnidirectional transmit antenna illuminating the scene and a spinning horn receive antenna collecting backscattered power as a function of azimuth. Average backscatter power ratio was found to be well modeled (under 3.3 dB RMS error) by a simple, previously- derived formula, parametrized by distance to a building across the street. Distribution of power variation in azimuth around this average is reproduced within 0.5 dB by a random azimuth spectrum with a lognormal amplitude distribution and random phase. Observed temporal backscatter power fluctuations were found to be well modeled by a Rician distribution, with K-factor values distributed lognormally. Observed clutter time fluctuation is an important effect expected to confound target detection as time variation is often regarded as a key property distinguishing moving targets from clutter, often assumed to be static. A more complete model of radar backscatter would include the delay dimension that most directly carries information on distance to target and other objects. One possible way to have that is to either collect wideband backscatter measurements to arrive at delay model or to rely on literature, as was done in [18] for indoor settings.

## IX. ACKNOWLEDGEMENTS

Thanks to Timothy Wang (University of Michigan), and Sergio Chapana (Pontificia Universidad Católica de Valparaíso) for assistance in measurements. A. Adhikari, J. Drogo and G. Zussman wish to acknowledge the support through NSF grants EEC- 2133516, CNS-2148128, AST-2232455, CNS-2450567. M. A. Almendra, M. Rodriguez, and R. Feick wish to acknowledge the support received from the Chilean Research Agency ANID, through research grants ANID FONDECYT 1250951, ANID CCTVal CIA250027, and ANID/Doctorado nacional/2023 under Grant 21232292.